\documentclass[traditabstract,print]{aa}
\usepackage{graphicx}
\usepackage{epsfig}
\usepackage{amssymb,amsmath}
\usepackage{natbib}
\bibpunct{(}{)}{;}{a}{}{,}
\usepackage{pslatex}
\usepackage{caption}
\usepackage{longtable}
\usepackage{hyperref}

\begin{document}

\title{Measuring the cosmic proper distance from fast radio bursts}

\author{H, Yu$^{1}$ and F. Y. Wang$^{1,2}$
\thanks{fayinwang@nju.edu.cn (FYW)}}

\institute{
$^{1}$ School of Astronomy and Space Science, Nanjing University, Nanjing 210093, China\\
$^{2}$ Key Laboratory of Modern Astronomy and Astrophysics (Nanjing
University), Ministry of Education, Nanjing 210093, China }


\authorrunning{Yu \& Wang}
\titlerunning{Measuring the cosmic proper distance from FRBs}


\abstract {The cosmic proper distance $d_P$ is a fundamental
distance in the Universe. Unlike the luminosity and angular diameter
distances, which correspond to the angular size, the proper distance
is the length of light path from the source to observer. However,
the proper distance has not been measured before. The recent
redshift measurement of a repeat fast radio burst (FRB) can shed
light on the proper distance. We show that the proper
distance-redshift relation can indeed be derived from dispersion measures
(DMs) of FRBs with measured redshifts. From Monte Carlo
simulations, we find that about 500 FRBs with DM and redshift
measurements can tightly constrain the proper distance-redshift
relation. We also show that the curvature of our Universe
can be constrained with a model-independent method using this
derived proper distance-redshift relation and the observed angular
diameter distances. Owing to the high event rate of FRBs, hundreds of
FRBs can be discovered in the future by upcoming instruments. The proper distance will play an important role in investigating the
accelerating expansion and the geometry of the Universe.}


\maketitle

\section{Introduction}\label{sec:intro}
In astronomy, a long-standing and intriguing question is the
measurement of distance. There are several distance definitions in
cosmology, such as the luminosity distance $d_L$, the angular diameter
distance $d_A$, the transverse comoving distance $d_M$ , and the
proper
distance $d_P$ \citep{Weinberg1972,Coles2002,Hogg1999}. In the frame
of the Friedmann-Lema\^{i}tre-Robertson-Walker (FLRW) metric, the proper
distance at the present time $t=t_0$, which is the same as the
comoving
distance, is \citep{Weinberg1972,Coles2002}
\begin{equation}\label{dP1}
    d_P(r) = a_0\int^r_0\frac{dr^\prime}{\sqrt{1-Kr^{\prime2}}}=a_0f(r),
\end{equation}
where $a_0$ is the present scale factor, $r$ is the comoving
coordinate of the source, and $f(r)$ is $\sin^{-1}r$, $r$, and
$\sinh^{-1}r$ for the curvature parameter $K=+1$, $K=0$, and $K=-1$,
respectively. Using the Hubble parameter $H=\dot{a}/a$, it can be
calculated from
\begin{equation}\label{dP2}
    d_P(z) = \frac{c}{H_0}\int^z_0\frac{dz^\prime}{E(z^\prime)},
\end{equation}
where $z$ is the redshift, $H_0$ is the Hubble constant, $c$ is the
speed of light, and $E(z)=H(z)/H_0$. Similarly, the transverse
comoving distance is \citep{Hogg1999}
\begin{equation}\label{dM}
    d_M(z)=a_0r(z)=\frac{c}{H_0\sqrt{-\Omega_K}}
    \sin[\sqrt{-\Omega_K}\int^z_0\frac{dz^\prime}{E(z^\prime)}],
\end{equation}
where $\Omega_K$ is the energy density fraction of cosmic curvature
($-i\sin(ix)=\sinh(x)$ if $\Omega_K>0$). The direct relation of
$d_M$, $d_A$ , and $d_L$ is $d_M=d_L/(1+z)=d_A(1+z)$.

\begin{figure*}
\centering
  \includegraphics[width=\textwidth]{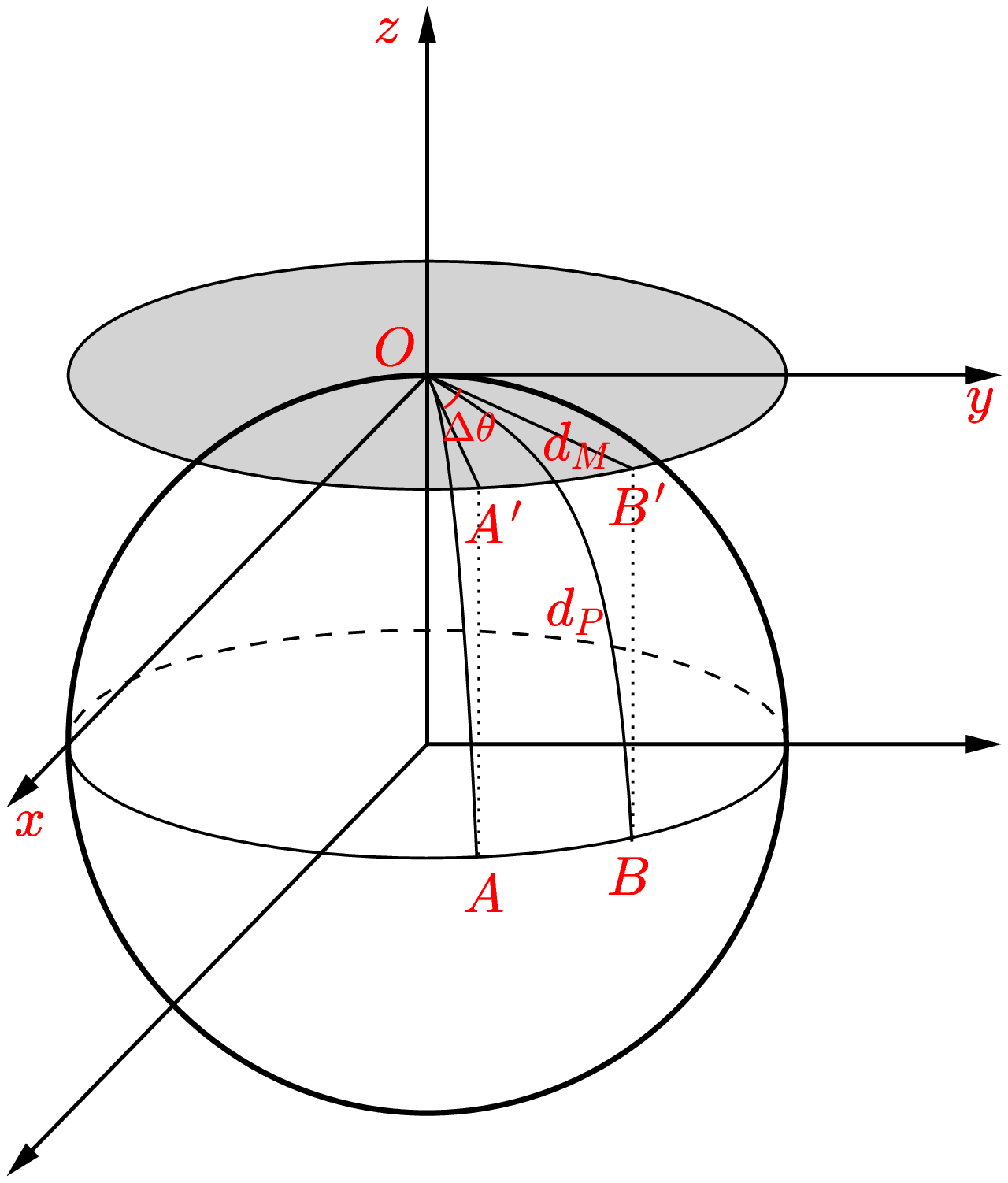}\\
  \caption{Illustration of the proper distance $d_P$ and transverse comoving distance $d_M$
  in a closed universe ($\Omega_K<0$). The source is $AB$, and observer is at $O$.
  It is obvious that the transverse comoving distance $d_M$ is shorter than the proper distance $d_P$. In a flat universe,
  they are the same, however. The cosmic curvature can therefore
be tested by comparing $d_P$ and $d_M$.}\label{fig1}
\end{figure*}

Many methods have been proposed to determine the cosmic distances.
For example, type Ia supernovae (SNe Ia), which are treated as
standard candles, have been used to measure the luminosity distance
$d_L$ \citep{Riess1998,Perlmutter1999}. The standard ruler (the
baryon
acoustic oscillation) has been used to derive the angular diameter
distance $d_A$ \citep{Eisenstein2005}. With the measurements of
$d_L$ and $d_A$, $d_M$ can be derived directly using the relation
among them. The luminosity and angular diameter distances have been
widely used in cosmology \citep[for recent reviews,
see][]{Weinberg2013,Wang2015}. Instead, the proper distance
$d_P$ is seldom used in cosmology because it is difficult to
measure \citep{Weinberg1972,Coles2002}. In a flat universe, the
transverse comoving distance $d_M$ and proper distance $d_P$ are
same. However, they are different in a curved universe. Figure
\ref{fig1} shows the differences between them in a closed universe.
In this figure, $AB$ is an object, and an observer at $O$ measures
the distance of $AB$. When the size of $AB$ and the angular size of
$\Delta\theta$ are known, the distance $d_A$, which is the length of
$OA^\prime$ or $OB^\prime$, can be determined. However, $d_M$ and
$d_L$ of $AB$ can also be derived using the relations among them.
The length of arc $OA$ or $OB$ is the physical
distance between the source and the observer, and it is the proper
distance $d_P$.

Whether our Universe is entirely flat is still unknown,
although the latest constraint on the cosmic curvature $|\Omega_K|$
is less than $0.005$ \citep{Planck2016}. However, we must keep in
mind that this constraint is model dependent, since it is derived in
$\Lambda$CDM background cosmology
\citep{Planck2016,Rasanen2015,Li2016}. Because of the differences
between the proper distance and other distances, three
important points encourage us to determine the proper distance
$d_P$. The first point is that the proper distance $d_P$ is the
fundamental distance in the Universe. The second point is that the
proper distance $d_P$ can be used to constrain the cosmic curvature
\citep{Yu2016}. The third point is to test the cosmological principle,
that is, the Universe is homogeneous and isotropic at large scales. The
basis of this idea is explained in Figure \ref{fig1}. The
transverse distance and proper distance of $AB$ can be regarded as
the lengths of the lines $OB^\prime$ and arc $OB$, respectively.
The ratios of the different parts of $OB^\prime$ and $OB$
can be used to test whether the curvatures at different scales of
the Universe are the same. In addition, the cosmological principle is expected to
be valid in the proper distance space rather than in $d_A$, $d_L$ , or
$d_M$ space if our Universe is not entirely flat. Therefore we should
test the cosmological principle in proper distance space unless we
can ensure that our Universe is entirely flat. With a similar
idea, R{\"a}s{\"a}nen et al. (2015) tested the FLRW metric using
the distance sum rule method \citep{Rasanen2015}. However, the
current
constraint obtained from this method is very loose.

In order to measure the proper distance, a
probe should in principle satisfy two conditions. First, it should change with redshift in a well-understood way and be independent of  cosmic
curvature. Second, it should record the information on the
expansion of our Universe. Standard candles and standard
rulers are not able to measure $d_P$ since they depend on
cosmic curvature. Up to now, no practical method to measure the
proper distance has been found. Fortunately, the discovery of fast
radio bursts (FRBs) \citep{Lorimer2007,Thornton2013} and their
redshifts \citep{Chatterjee2017,Tendulkar2017} sheds light on
deriving the proper distance-redshift relation. A radio signal
traveling through plasma exhibits a quadratic shift in its arrival
time as a function of frequency, which is known as the dispersion
measure (DM). The DM of radio signal is proportional to the
integrated column density of free electrons along the line of sight
(i.e., DM$\propto\int n_edl$), which was widely used in Galactic
pulsar data \citep{Taylor1993,Manchester2005} and gamma-ray bursts
\citep{Ioka2003,Inoue2004}. In addition, the redshift measurement of
the source gives information on the expansion of the Universe.
For an FRB, the DM can be measured directly
\citep{Lorimer2007,Thornton2013}, which has been proposed for
cosmological purposes \citep{Zhou2014,Gao2014,Lorimer2016}.
Its redshift can be estimated by observing its
host galaxy or afterglow \citep{Lorimer2016,Tendulkar2017}.
Therefore, the $d_p-z$ relation can be derived with the DM and
redshift measurements of a large FRB sample.

This paper is organized as follows. In Sect.
\ref{sec:method} we introduce the method used to determine the
$d_P-z$ relation. In Sect. \ref{sec:simulation} we use a Monte
Carlo simulation to test the validity and efficiency of our method.
We summarize our result in Sect.
\ref{sec:discussion}.

\section{Method for determining the $d_P-z$ relation}\label{sec:method}
\subsection{Main idea}
FRBs are millisecond-duration radio signals occurring at
cosmological distances \citep{Tendulkar2017}. The DM of FRB caused by the intergalactic
medium (IGM) is
\begin{equation}\label{DM_igm1}
    {\rm DM_{IGM}} = \Omega_b\frac{3H_0c}{8\pi
    Gm_p}\int^z_0\frac{F(z)}{E(z^\prime)}dz^\prime,
\end{equation}
where
\begin{eqnarray*}
  F(z) &=& (1+z)f_{IGM}(z)f_e(z).
\end{eqnarray*}
$\Omega_b$ is the baryon mass density fraction of the universe, $G$
is the gravitational constant, $m_p$ is the rest mass of protons,
$f_{\rm IGM}$ is the fraction of baryon mass in the intergalactic
medium (IGM), $f_e =
Y_HX_{e,H}(z)+\frac{1}{2}Y_{He}X_{e,He}(z)$ represents the average
count of electrons contributed by each baryon, $Y_H=3/4$
and
$Y_{He}=1/4$ are the mass fractions of hydrogen and helium, and
$X_{e,H}$ and $X_{e,He}$ are the ionization fractions of intergalactic
hydrogen and helium, respectively. The parameters in $F(z)$ are
extensively investigated in previous works
\citep{Fan2006,McQuinn2009,Meiksin2009,Becker2011}. According to
their results, intergalactic hydrogen and helium
are fully ionized at $z<3$. Therefore we chose $X_{e,H}=X_{e,He}=1$ at $z<3,$ which
corresponds to $f_e=0.875$. The values of $f_{\rm IGM}$ are $0.82$
and $0.9$ at $z<0.4$ and $z>1.5$, respectively \citep{Meiksin2009,Shull2012}. To describe the slow
evolving of $f_{\rm IGM}$ in the range $0.4<z<1.5$, we assumed that it
increases linearly at $0.4<z<1.5$ \citep{Zhou2014}. If
FRBs can be detected in a wide range of redshifts, we can therefore
use the
observed ${\rm DM_{IGM}}-z$ relation to determine the $d_P-z$
relation by removing the effect of $F(z)$.

When an FRB signal travels through the plasma from the source to
the observer, its DM can be measured with high accuracy. However, this
$\rm DM_{obs}$ includes several components that are caused by the
plasma in the IGM, the Milky Way, the host galaxy of the FRB, and even
the source itself. It has
\begin{equation}\label{DM_igm2}
\rm DM_{IGM} = DM_{obs}-DM_{MW}-\frac{DM_{host}+DM_{source}}{1+z}.
\end{equation}
Only the $\rm DM_{IGM}$ contains the information of the proper distance.
Other components therefore need to be subtracted from the $\rm
DM_{obs}$. Since the $\rm DM_{MW}$ is well understood through pulsar
data \citep{Taylor1993,Manchester2005}, it can be subtracted.
Alternatively, we can only use those FRBs at high galactic
latitude that have low $\rm DM_{MW}$ \citep{Zhou2014,Gao2014}. For
the local $\rm DM_{loc}$, which contains the $\rm DM_{host}$ and
$\rm DM_{source}$, the recent finding of the host galaxy of the
repeating FRB 121102 suggests a low value $\lesssim324 \rm
pc\,cm^{-3}$ and it is probably even lower depending on geometrical
factors \citep{Tendulkar2017}. The variation in total DM for FRB
121102 is very small \citep{Spitler2016,Chatterjee2017}, which
indicates that the $\rm DM_{loc}$ is almost constant. Moreover,
\cite{Yang2016} proposed a method to determine it based on the
assumption that $\rm DM_{loc}$ does not evolve with redshift. More
fortunately, the $\rm DM_{loc}$ should be decreased by dividing a
$1+z$ factor since the cosmological time delay and frequency shift.
While $\rm DM_{IGM}$ increases with redshift,  $\rm
DM_{loc}$ is not important at high redshifts. Therefore we
can subtract the $\rm DM_{loc}$ from $\rm DM_{obs}$ and leave its
uncertainty into the total uncertainty $\sigma_{\rm tot}$ which is
the uncertainty of $\rm DM_{IGM}$ extracted from $\rm DM_{obs}$. It
has
\begin{equation}\label{sigma_DM}
\sigma^2_{\rm tot}=\sigma^2_{\rm obs}+\sigma^2_{\rm
MW}+\frac{\sigma^2_{\rm DM_{loc}}}{(1+z)^2}+\sigma^2_{{\rm
DM_{IGM}}(z)}.
\end{equation}
Since the accurate measurement of DM and the well-understood
measurement of $\rm
DM_{GW}$, $\sigma_{\rm obs}$ , and $\sigma_{\rm MW}$ can be omitted
compared with the much larger $\rm \sigma_{DM_{loc}}$ and $\rm
\sigma_{DM_{IGM}}$. Following \cite{Thornton2013} and
numerical simulations of \cite{McQuinn2014}, we chose $\rm
\sigma_{DM_{loc}}=100\,pc/cm^3$ and $\rm
\sigma_{DM_{IGM}}=200\,pc/cm^3$ in the following analysis. These
uncertainties are nuisance parameters in an analysis. Fortunately, they can be
decreased by using the average $\rm DM_{IGM}$ when there are tens of
FRBs in a narrow redshift bin \citep{Zhou2014} (for example, $\Delta
z\sim0.06$).

Recently, the host galaxy of FRB 121102 was identified, which can
give accurate redshift information
\citep{Chatterjee2017,Tendulkar2017}. When enough FRBs with redshifts
are observed, the $d_P-z$ relation can be derived from the ${\rm
DM_{IGM}}-z$ relation. Based on the high FRB rate $10^4\rm
sky^{-1}\,day^{-1}$ \citep{Thornton2013}, a large sample of FRBs
may be collected in the future, which will become the basis of FRB
cosmology.

The DM$_{\rm IGM}$ contains the $d_P$ information and mixes it with
$F(z)$, which corresponds to the anisotropic distribution of free
electrons in the Universe. When the effect of $F(z)$ is removed, the
$d_P-z$ relation can be derived from the DM$_{\rm IGM}-z$ relation.
The first step therefore is to reconstruct the $\rm DM_{IGM}(z)$
function of the FRB. The Gaussian process (GP) is a model-independent method
to solve this type of problem. The advantage of the GP is that it
can reconstruct a function from data without assuming any function
form (for more details about the GP, see the next subsection and
\cite{GP_ref}). With the GP method, we can therefore obtain the
DM$_{\rm IGM}-z$ relation from FRB observational data without any
cosmological model assumption. Then we remove the effect of $F(z)$
to obtain the model-independent $d_P-z$ relation. In this work, we
used the python code package GaPP developed by \cite{Seikel2012}. GaPP can reconstruct the function of given data as well as its
first, second, and third derivative functions (see \cite{Seikel2012}
for more details about GaPP).

With the data set ($z$, ${\rm DM_{obs}}$) of a future sample of
observed FRBs, we can use the steps as follows to derive the
$d_P(z)$.
\begin{itemize}
  \item Subtracting the $\rm DM_{MW}$ and $\rm DM_{loc}$ to obtain the data set ($z$, ${\rm DM_{IGM}}$).
  \item Dividing the data set ($z$, ${\rm DM_{IGM}}$) into several redshift bins, each of which contains tens of FRBs, and then calculating the average redshift and $\rm DM_{IGM}$ and the standard deviation of $\rm DM_{IGM}$. Then we have a data set $\langle z\rangle$, ${\rm \langle DM_{IGM}\rangle}$, and $\sigma_{\rm \langle DM_{IGM}\rangle}$.
  \item Using the GP method to reconstruct the function ${\rm DM_{IGM}}(z)$ and its first derivative function $G(z) = {\rm d}{\rm DM_{IGM}}(z)/{\rm d}z$.
  \item Reintegrating the function $I(z)=\frac{G(z)}{AF(z)}$, which should be $c/H(z)$, to obtain $d_P(z),$ where $A=\Omega_bH_0^2\frac{3c}{8\pi Gm_p}$ , and $\Omega_bH_0^2$ can be given by other observations.
\end{itemize}

\subsection{Gaussian process}
The GP is a statistical model to smooth a
continued function from discrete data. In this model, the value of
the function $f(x)$ at any point $x$ is assumed to be a random variable
with normal distribution. The mean and Gaussian error value,
$\mu(x)$ and $\sigma(x)$, are determined by all of the observed data
through a covariance function (or kernel function) $K(x,\tilde{x})$,
$f(\tilde{x}),$ and $\sigma_{\tilde{x}}$ , where $\tilde{x}$s are the
points with observed data and $\sigma_{\tilde{x}}$s are their
errors. It has
\begin{equation}\label{mu}
  \mu(x) = K(x,\tilde{x})(K(\tilde{x},\tilde{x})+\sigma_{\tilde{x}}^2I)^{-1}f(\tilde{x})
\end{equation}
and
\begin{equation}\label{sigma}
  \sigma(x) = K(x,x)-K(x,\tilde{x})(K(\tilde{x},\tilde{x})+\sigma_{\tilde{x}}^2I)^{-1}K(\tilde{x},x).
\end{equation}
When the kernel function is given, we can use the GP to derive
the distribution of the continued function $f(x)$.

As we described above, we used the open-source Python package Gapp to
apply the GP. This code is
widely used \citep{Cai2016,Yu2016}. The kernel function in this
code is
\begin{equation}\label{kernel}
  K(x,x^\prime)=\sigma_f^2\exp(-\frac{(x-x^\prime)^2}{2l^2}),
\end{equation}
where $\sigma_f$ and $l$ are two parameters to describe the
amplitude and length of the correlation in the function value and $x$
directions, respectively. These two parameters can be optimized by the GP
with the observational data $f(\tilde{x})$ through maximizing their
log marginal likelihood function \citep{Seikel2012},
\begin{eqnarray}\label{likelihood}
  \ln{\mathcal{L}} &=& \ln{p(f(\tilde{x})|\tilde{x},\sigma_f,l)} \\
  \nonumber
   &=&
   -\frac{1}{2}(f(\tilde{x})-\mu(\tilde{x}))^T[K(\tilde{x},\tilde{x})+\sigma_{\tilde{x}}^2I]^{-1}(f(\tilde{x})-\mu{\tilde{x}})\\ \nonumber
   &&- \frac{1}{2}\ln{|K(\tilde{x},\tilde{x})+\sigma_{\tilde{x}}^2I|}-\frac{N}{2}\ln{2\pi,}
\end{eqnarray}
where $N$ is the number of observed data. In the Gapp package, all
of this can be calculated automatically.

\section{Simulations and results}\label{sec:simulation}
We tested the efficiency of our method using Monte Carlo simulations.
First we used Eqs. (\ref{DM_igm1}) and (\ref{sigma_DM}) to
create a mock data set ($z$, ${\rm DM_{IGM}}$, $\sigma_{\rm tot}$)
under a background cosmology. Then we used the above method to derive
the $d_P(z)$ function and compare it with theoretical $d_P(z)$. A
flat $\Lambda$CDM cosmology with parameters $\Omega_b=0.049$,
$\Omega_m=0.308$, $\Omega_\Lambda=1-\Omega_m$ , and $H_0=67.8$
km/s/Mpc was assumed \citep{Planck2016}. The redshift distribution of
the FRBs was assumed as $f(z)\propto ze^{-z}$ in the redshift range
$0<z<3$, which is similar as the redshift distribution of long gamma-ray bursts
\citep{Zhou2014,Shao2011}. In order to avoid random uncertainty,
we simulated $10^4$ times. In each simulation are 500 mock
$\rm DM_{IGM}$, which are equally separated into 50 bins in redshift
space.

\begin{figure*}
\centering
  \includegraphics[width=\textwidth]{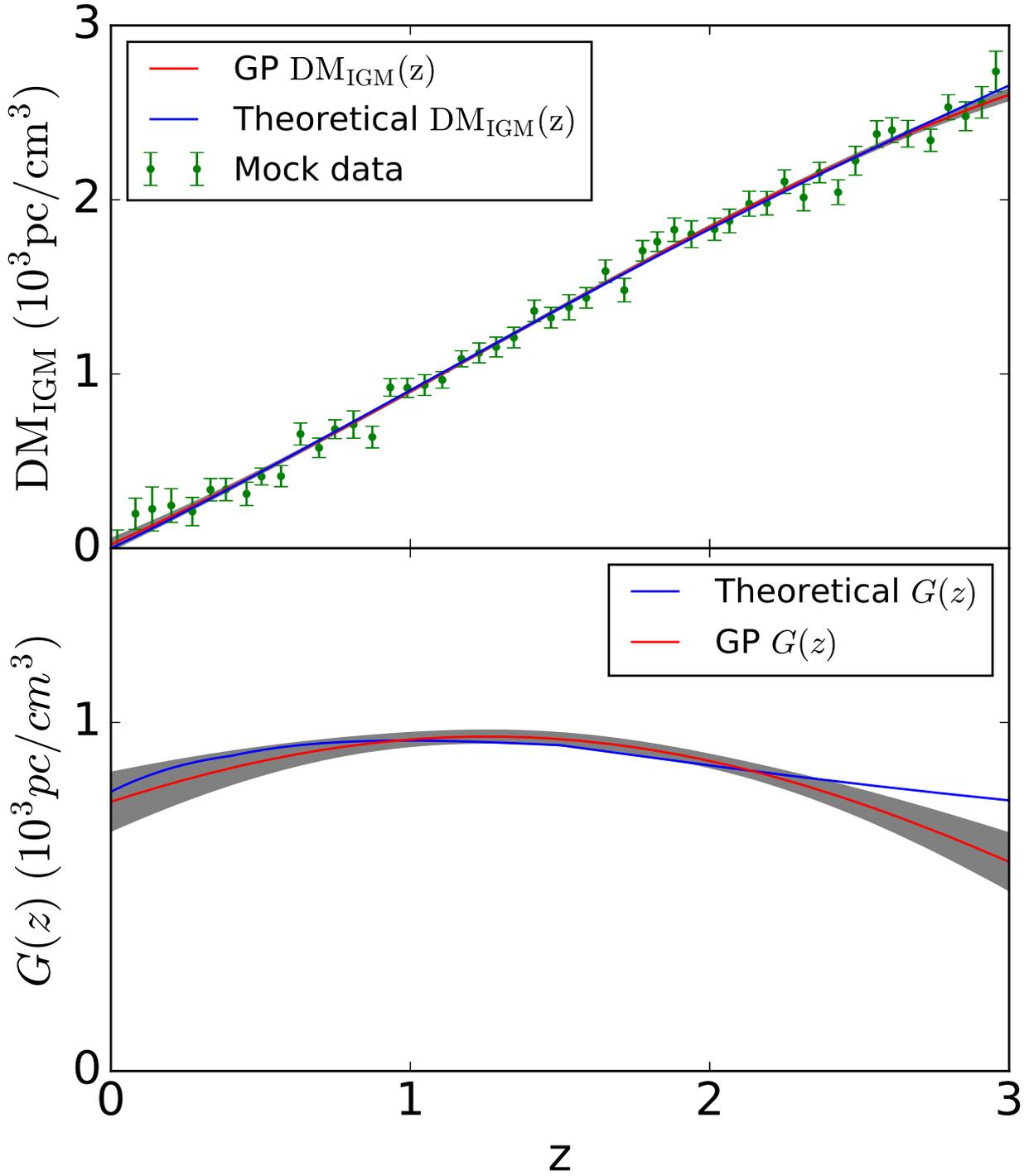}\\
  \caption{Top panel: binned mock $\rm DM_{IGM}$ data with $1\sigma$ errors, the GP reconstructed ${\rm DM_{IGM}}(z)$ function with its
$1\sigma$
   confidence region, and its theoretical function. Bottom panel: $G(z)$ function with its $1\sigma$
   confidence region derived from the GP method and its theoretical function.
   $\Omega_K=0$ is assumed.}\label{fig2}
\end{figure*}

\begin{figure*}
\centering
  \includegraphics[width=\textwidth]{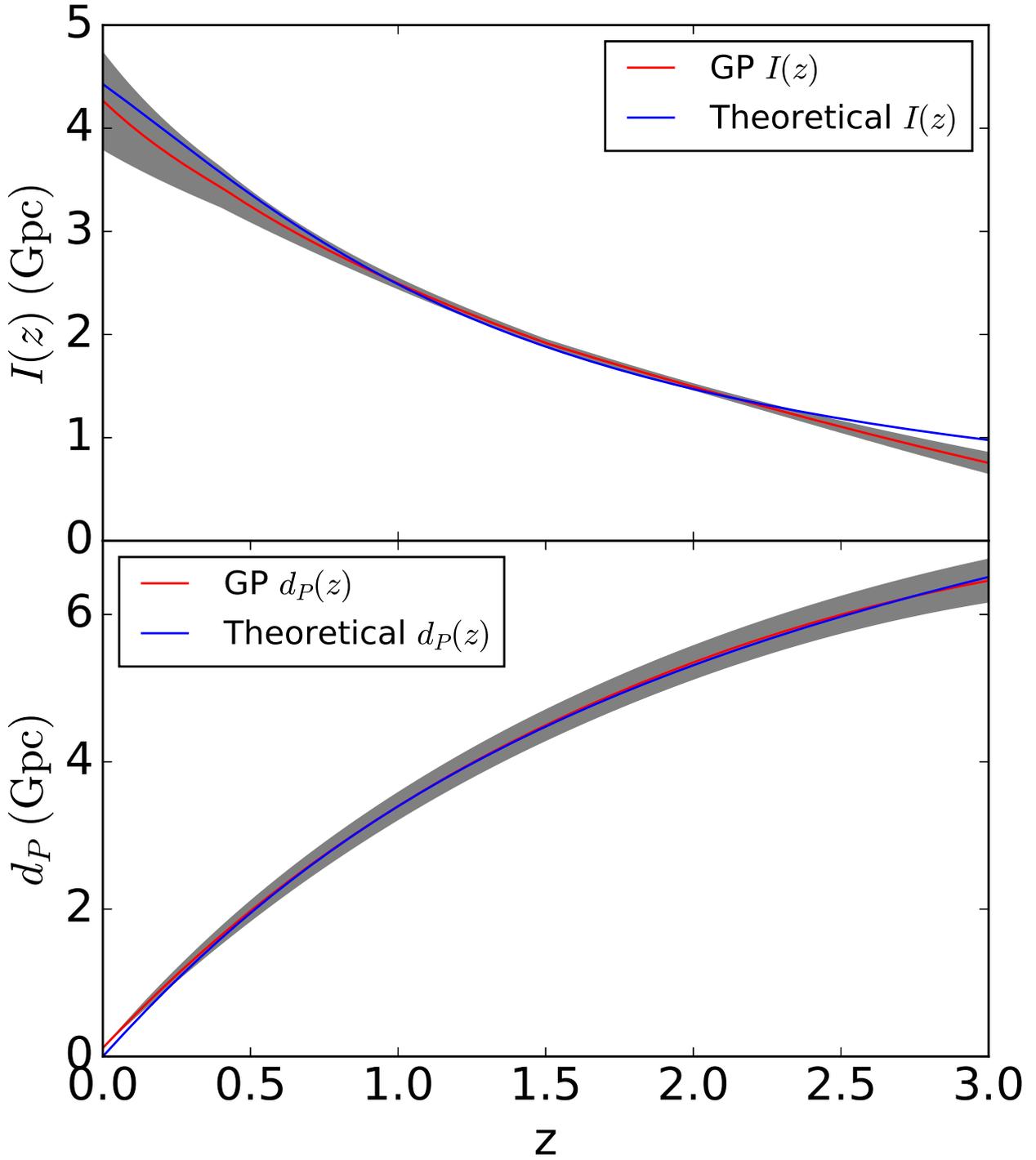}\\
  \caption{Top panel: $I(z)$ function with its $1\sigma$
   confidence region derived from the GP method and its theoretical function. Bottom panel: same as the top panel, but for the derived $d_P(z)$ function.
   $\Omega_K=0$ is assumed.}\label{fig3}
\end{figure*}

Figures \ref{fig2} and \ref{fig3} show an example of $10^4$
simulations in the $\Omega_K=0$ case. The top panel of Figure
\ref{fig2} shows the binned DM$_{\rm IGM}$ data (green dots), the
reconstructed $\rm{DM_{IGM}}(z)$ function derived with the GP method
(red line), and the theoretical function (blue line). The bottom panel
gives the derived and theoretical function $G(z)$. Similar as Figure
\ref{fig2}, Figure \ref{fig3} shows the derived and theoretical
$I(z)$ and $d_P(z)$ function. These figures show that
the reconstructed $\rm{DM_{IGM}}(z)$ function and the final derived
$d_P(z)$ function are well consistent with theoretical functions,
although the $G(z)$ and $I(z)$ functions in middle steps are slightly
biased. This shows that the $d_P(z)$ function derived from mock $\rm
DM_{IGM}$ data with the GP method is reliable.

\begin{figure*}
\centering
  \includegraphics[width=\textwidth]{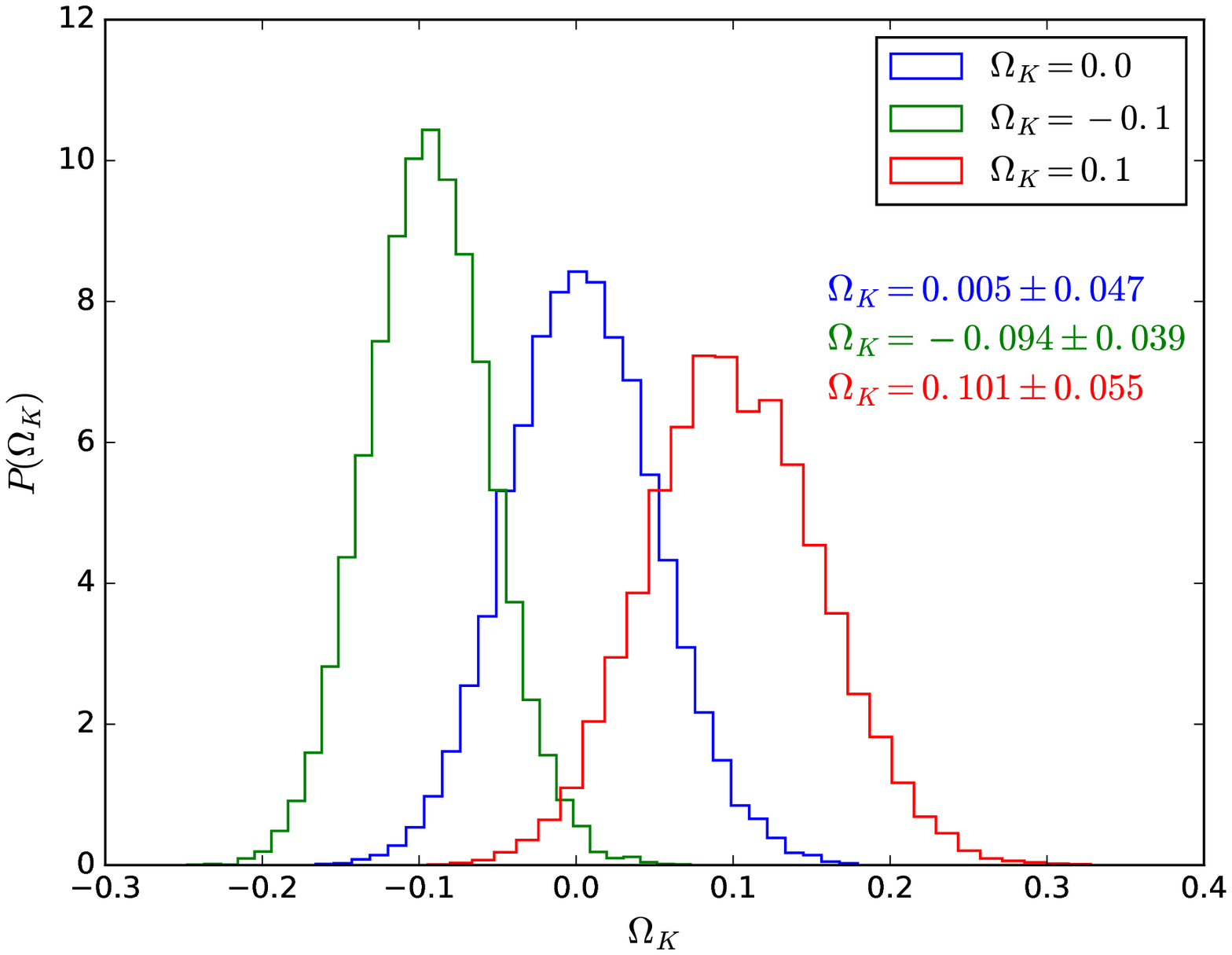}\\
  \includegraphics[width=0.4\textwidth]{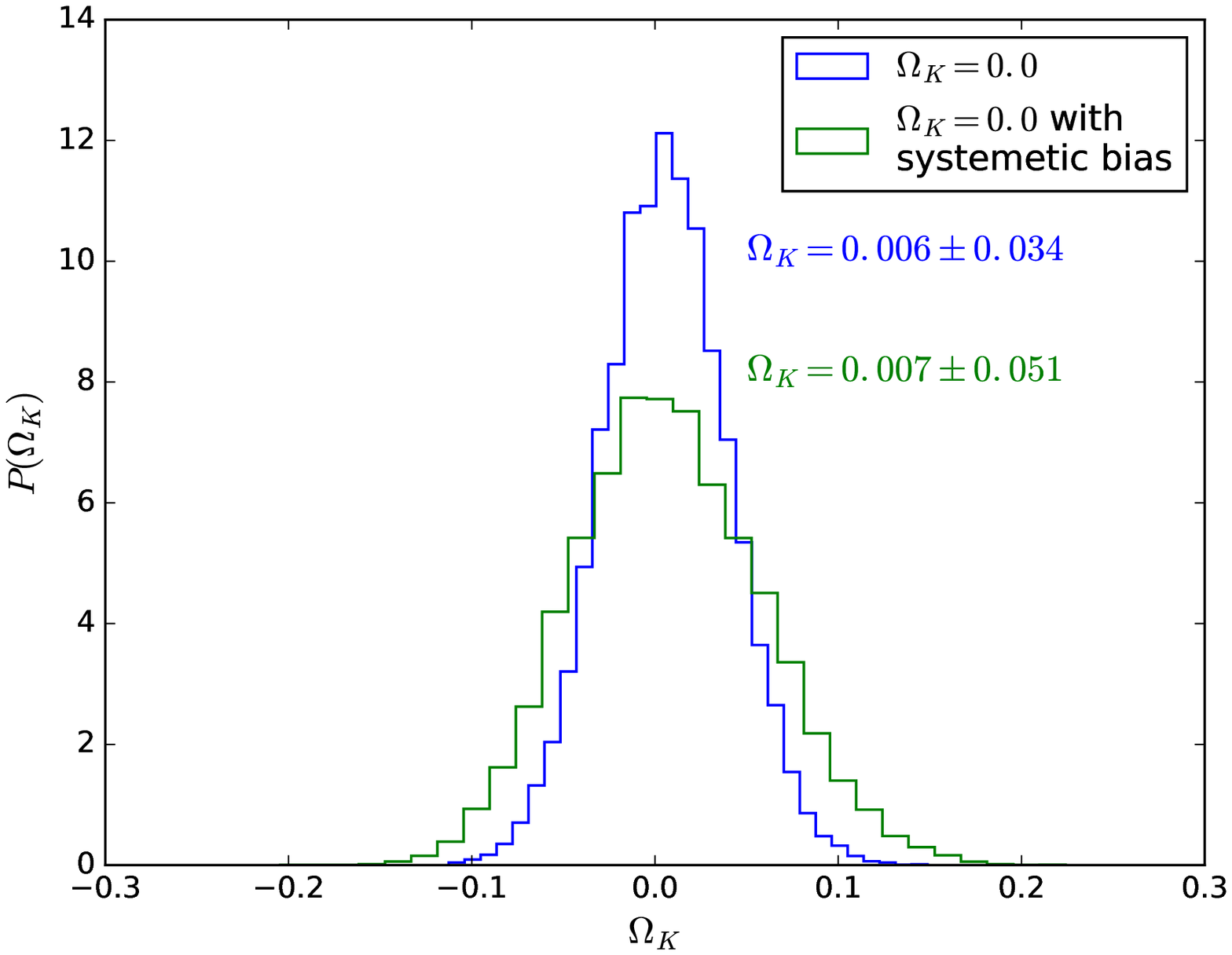}
  \includegraphics[width=0.4\textwidth]{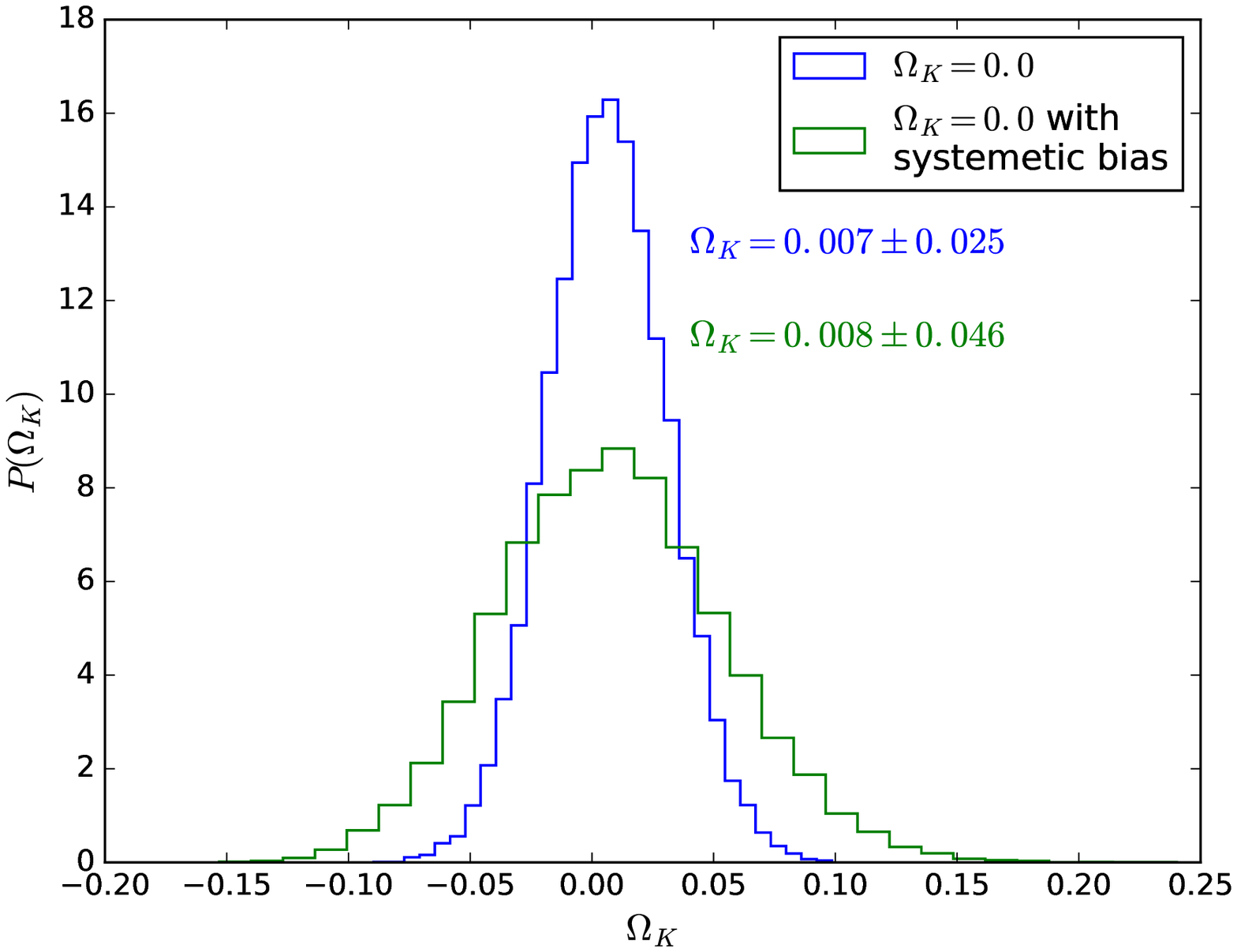}
  \caption{Top panel: posterior distributions of $\Omega_K$ in three different
   $\Omega_K$ cases with 500 mock FRBs data. The derived $\Omega_K$ value are clearly well
   consistent with the assumed values. The bottom two panels show the ability of our method when the sample includes 1000 and 2000 FRBs.
   The blue and green histograms show the results without and with the systematic uncertainty, respectively.}\label{fig4}
\end{figure*}

We also tested the validity and efficiency of our method using the
equation
\begin{equation}\label{dpdm}
\frac{H_0d_M}{c}\sqrt{-\Omega_K} =
\sin(\frac{H_0d_P}{c}\sqrt{-\Omega_K}),
\end{equation}
which can constrain $\Omega_K$ independently of the model \citep{Yu2016}.
First, 20 mock transverse comoving distance $d_M$ data were uniformly
simulated
from Eq. (\ref{dM}) in the redshift range $1.0<z<3.0.$
 Then we performed the same simulations as introduced above in
three different fiducial $\Omega_K$ cases,  -0.1, 0, and 0.1.
Next, we compared the simulated $d_M$ data with the $d_P(z)$ function
derived with the GP method and used Eq. (\ref{dpdm}) to solve
$\Omega_K$. Finally, we took the average value of them and compared
it with the fiducial value. To avoid the randomness of simulation,
we also simulated this $10^4$ times for each case and drew the posterior
probability distributions of the mean $\Omega_K$. The top panel of
the Figure \ref{fig4} shows the posterior probability distributions
of $\Omega_K$ in three different fiducial $\Omega_K$ cases,
-0.1, 0, and 0.1. The assumed cosmic curvatures can be well recovered
with errors $\sigma\approx0.05$ using 500 FRBs data, which also
means that the $d_P(z)$ function derived from mock $\rm DM_{IGM}$
data with the GP method is reliable and can be used to constrain the
cosmic curvature. The bottom two panels show that the errors will
decrease to $\sigma\approx0.034$ and 0.025 when the FRB sample
contains 1000 and 2000 FRBs, respectively (the blue histograms in
the bottom panels of Figure \ref{fig4}).

When deriving the function $d_P(z)$, it must be noted that the prior of
$\Omega_bh_0^2$ with $h_0=H_0/100\rm\,km/s/Mpc$, and the function
$F(z)$ in Eq. (\ref{DM_igm1}) will introduce some uncertainties
into the derived $d_P(z)$ function. For the $F(z)$ function, which
describes the distribution of free electrons in the Universe, we can
include its contribution to the $\sigma_{\rm DM_{IGM}}$. We chose $\sigma_{\rm DM_{IGM}}=200\rm\,pc/cm^{3}$  here, which
includes the potential effects of the uncertainty of the function
$F(z)$. The more important and nuisance point is the systematic
uncertainty caused by the choice of the prior of $\Omega_bh_0^2$.
From the expression of $I(z)$, it is easy to find that the value of
$\Omega_bh_0^2$ will directly affect the derived $d_P(z).$  In order
to evaluate the effect of $\Omega_bh_0^2$, we considered a Gaussian
uncertainty for $\Omega_bh_0^2$ and repeated the Monte Carlo
simulations. Since the uncertainty of $\Omega_bh_0^2$ is
about 1\% \citep{Planck2016}, we chose the value of the systematic
uncertainty as 1\%. The results are shown as the green histograms in
the bottom panels of Figure \ref{fig4}. For 1000 FRBs with
redshift measurements, the uncertainty of the $\Omega_K$ is
about 0.05, which is acceptable. However, the exact value of $H_0$
is still unknown. For the value of $H_0$, the
value from Cepheid+SNe Ia and the cosmic microwave background (CMB) differs. For
example, Riess et al. (2016) derived the best estimate of
$H_0=73.24\pm1.74$~km~s$^{-1}$~Mpc$^{-1}$ using Cepheids, which is
about 3.4$\sigma$ higher than the value from \cite{Planck2016}.
However, Aubourg et al. (2015) used the 2013 Planck data in
combination with BAO and the JLA SNe data to find
$H_0=67.3\pm1.1$~km~s$^{-1}$~Mpc$^{-1}$, in excellent agreement with
the 2015 Planck value. Moreover,
$H_0=62.3\pm6.3$~km~s$^{-1}$~Mpc$^{-1}$ is derived from the
Cepheid-calibrated luminosity of SNe Ia \citep{Sandage06}, which
agrees with the 2015 Planck value. Therefore we used the best
constraint on $\Omega_bh_0^2$ from Planck CMB data.

The systematic uncertainty is always a nuisance problem in
all cosmology probes, such as SNe Ia as the probe of the luminosity
distances and BAO as the probe of the angular diameter distances.
SNe Ia are widely accepted to be excellent standard candles at
optical wavelengths. The luminosity distances could be derived from
SNe Ia. However, the exact nature of the binary progenitor system (a
single white dwarf accreting mass from a companion, or the merger of
two white dwarfs) still is an open question. Systematic
errors, including calibration, Malmquist bias, K-correction, and dust
extinction, degrade the quality of SNe Ia as standard candles
\citep{Riess2004}. Moreover, the derived luminosity distances are
not fully model independent \citep{Suzuki2012}. The derived
cosmological parameters from SNe Ia are significantly biased by
systematic errors (see Figure 5 of Suzuki et al. (2012) for details).
The angular diameter distances for galaxy clusters can be obtained
by combining the Sunyaev-Zeldovich temperature decrements and X-ray
surface brightness observations. The error of the angular diameter
distance can be up to 20\% \citep{Bonamente2006}, however. Therefore the
$d_P(z)$ derived from FRBs can be a supplementary tool,
although greater efforts are required on its systematic error.

\section{Summary}\label{sec:discussion}
In cosmology, the proper distance $d_P$ corresponds to the length of
the light path between two objects. It is a potentially useful tool to
test the cosmic curvature and cosmological principle. In the past,
the proper distance was seldom used to investigate our Universe since
it is difficult to measure. We proposed a model-independent method
to derive the proper distance-redshift relation $d_P(z)$ from DM and
redshift measurements of FRBs. The basis of our method is that many
FRBs with measured redshifts and DMs may be observed in a wide
redshift range (i.e., $0<z<3$) in the future. This is possible
because of the high rate of FRBs, which is about $10^4~\rm
sky^{-1}\,day^{-1}$ \citep{Thornton2013}. Although some authors have
used FRBs as cosmological probes \citep{Zhou2014,Gao2014}, they only
considered DMs. The most important point is that the distance
information contained in DMs is the proper distance $d_P$, whose
difference with other distances is important in understanding the
fundamental properties of our Universe.

In the near future, several facilities such as the Canadian Hydrogen
Intensity Mapping Experiment (CHIME) radio telescope,
the Five-hundred-meter Aperture Spherical Telescope (FAST) in China, and
the Square Kilometer Array will commence working. Interestingly, the CHIME
might detect dozens of FRBs per day\citep{Kaspi2016}. A
large sample of FRBs with redshift measurements is therefore expected in next
decade \citep{Lorimer2016}. With a large sample of FRBs, the proper
distance derived from FRBs will be a new powerful cosmological
probe.

\section*{Acknowledgements}\label{sec:ack}
We thank the anonymous referee for constructive comments. This work
is supported by the National Basic Research Program of China (973
Program, grant No. 2014CB845800), the National Natural Science
Foundation of China (grants 11422325 and 11373022), the Excellent
Youth Foundation of Jiangsu Province (BK20140016). H. Yu is also
supported by the Nanjing University Innovation and Creative Program for
PhD candidates (2016012).

\newpage

\clearpage

\clearpage

\clearpage

\end{document}